\documentclass[review]{elsarticle}

\usepackage{mathtools}
\usepackage{lineno,hyperref}
\usepackage{cleveref}
\usepackage{color,xspace}
\modulolinenumbers[5]

\journal{Nuclear Instruments and Methods in Physics Research A }

\newcommand{\mue}{$\mu$$-$$e$\xspace}

\newcommand{\gasisobutane}{He$-$iC$_{4}$H$_{10}$(90/10)\xspace}

\newcommand{\dedx}{d$E$/d$x$ }
\newcommand{\strel}{$s$-$t$ }

\begin{document}

\begin{frontmatter}

\title{Test of a small prototype of the COMET cylindrical drift chamber}

\author[osaka,nanjing]{C.~Wu$^{1,}$}   
\author[osaka]{T.S.~Wong} 
\author[osaka]{Y.~Kuno$^{1,}$}   
\author[osaka,kek,ihep-epd]{M.~Moritsu} 
\author[osaka]{Y.~Nakazawa$^{2,}$}   
\author[osaka]{A.~Sato} 
\author[osaka]{H.~Sakamoto$^{3,}$}  
\author[osaka]{N.H.~Tran$^{4,}$}   
\author[osaka]{M.L.~Wong$^{5,}$}  
\author[osaka]{H.~Yoshida$^{1,}$}   
\author[osaka]{T.~Yamane} 
\author[ihep-skl,ucas]{J.~Zhang} 

\address[osaka]{Department of Physics, Osaka University, Toyonaka, Osaka 560-0043, Japan}
\address[nanjing]{Department of Physics, Nanjing University, Nanjing 210093, China}
\address[kek]{Institute of Particle and Nuclear Studies, High Energy Accelerator Research Organization, Tsukuba 305-0801, Japan}
\address[ihep-epd]{Experimental Physics Division, Institute of High Energy Physics, Chinese Academy of Sciences, Beijing 100049, China}
\address[ihep-skl]{State Key Laboratory of Particle Detection and Electronics, Institute of High Energy Physics, Chinese Academy of Sciences, Beijing 100049, China}
\address[ucas]{University of Chinese Academy of Sciences, Beijing 100049, China}

\fntext[fn1]{Present address: Research Center for Nuclear Physics, Osaka University, Ibaraki, Osaka 567-0047, Japan}
\fntext[fn2]{Present address: Institute of Particle and Nuclear Studies, High Energy Accelerator Research Organization, Tsukuba 305-0801, Japan}
\fntext[fn3]{Present address: Centre for Computational Science, RIKEN, Kobe, Japan}
\fntext[fn4]{Present address: Department of Physics, Boston University, Boston, MA 02215, USA}
\fntext[fn5]{Present address: Laboratoire de Physique de Clermont, Campus Universitaire des C\'{e}zeaux 4 Avenue Blaise Pascal 63178 Aubi\`{e}re Cedex, France}




\begin{abstract}

The performance of a small prototype of a cylindrical drift chamber (CDC) used in the COMET Phase-I experiment was studied by using an electron beam. 
The prototype chamber was constructed with alternating all-stereo wire configuration and operated with the \gasisobutane gas mixture without a magnetic field. 
The drift space-time relation, drift velocity, \dedx resolution, hit efficiency, and spatial resolution as a function of distance from the wire were investigated. 
The average spatial resolution of $150\,\mu$m with the hit efficiency of $99\,\%$ was obtained at applied voltages higher than 1800\,V. 
We have demonstrated that the design and gas mixture of the prototype match the operation of the COMET CDC. 

\end{abstract}

\begin{keyword}
  \texttt{Drift chamber}\sep
  \texttt{helium-based gas}\sep
  \texttt{spatial resolution}\sep
  \texttt{COMET}
\end{keyword}

\end{frontmatter}


\section{Introduction}

The COherent Muon to Electron Transition (COMET) experiment at the Japan Proton Accelerator Research Complex (J-PARC) in Tokai, Japan, is to search for the neutrinoless coherent transition of a muon to an electron (\mue conversion) in the field of a nucleus, with its single event sensitivity of $2\times 10^{-17}$, which is more than four orders of magnitude improvement over the current upper limit of $7 \times 10^{-13}$ at 90\,\% C.L. from the SINDRUM II experiment \cite{Bertl:2006up}. 
The COMET experiment is organized as a two-stage project. 
The first stage, COMET Phase-I, is seeking a sensitivity of $3 \times 10^{-15}$ with 3.2-kW proton beam power \cite{Kuno:2013mha,Adamov:2018vin}. 
It will be carried out at the Hadron Experimental Facility of J-PARC using a bunched 8-GeV proton beam that is slowly extracted from the J-PARC Main Ring.

\begin{figure}[htb!]
 \begin{center}
 \includegraphics[width=1.0\textwidth]{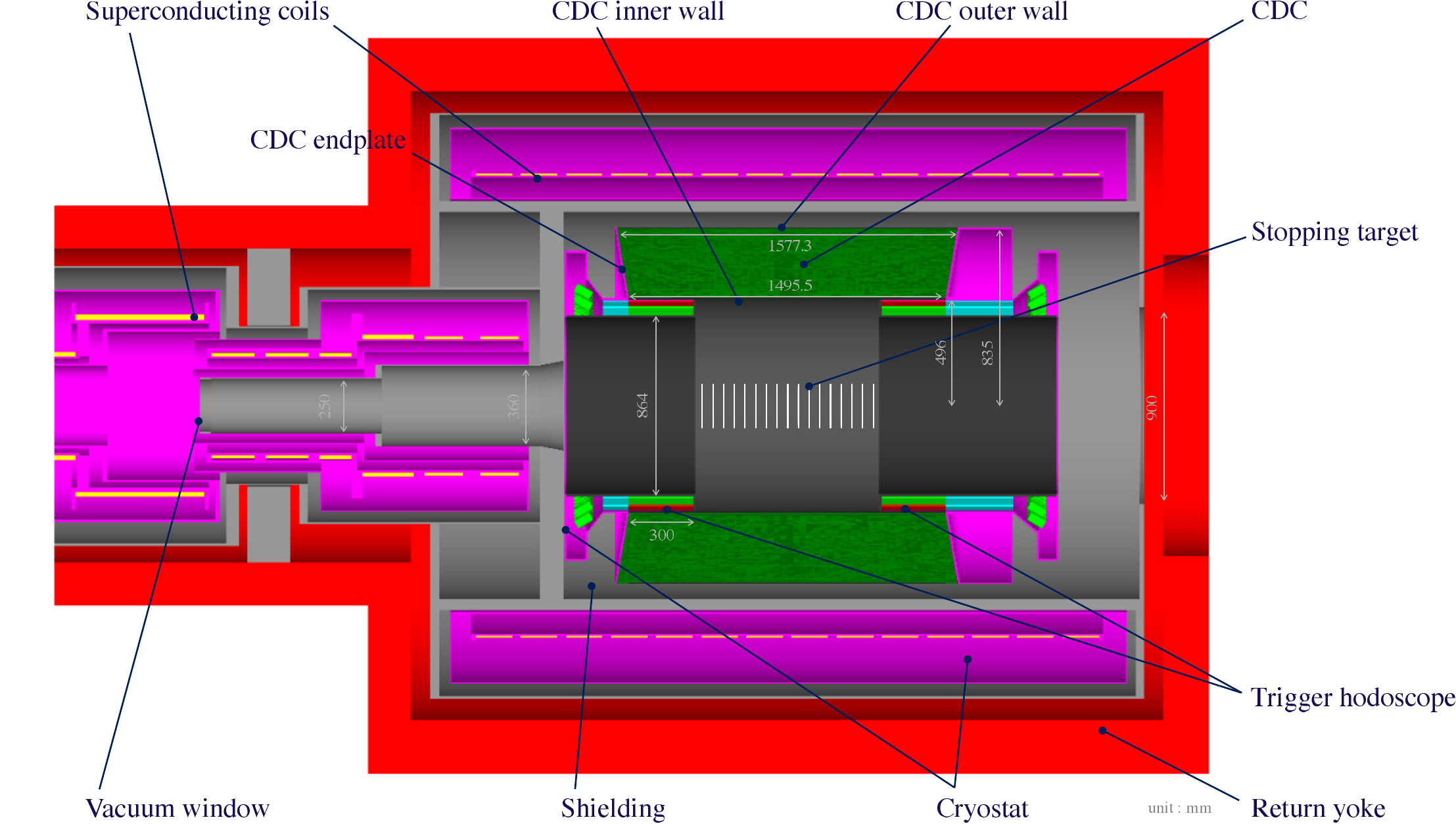}
 \end{center}
 \caption{\sl Schematic layout of the cylindrical detector system. A muon beam comes from left and stops in the target disks located in the center of the detector solenoid.}
 \label{fig:cydet}
\end{figure}

The main detector system for the \mue conversion search in COMET Phase-I is the Cylindrical Detector (CyDet). 
It consists of a cylindrical drift chamber (CDC) and a cylindrical trigger hodoscope. 
The primary function of the CDC is to measure the momentum of \mue conversion electrons. 
\Cref{fig:cydet} shows a schematic layout of the CyDet. 
It is located downstream of the muon transport section, and installed within the warm bore of a large 1-T superconducting detector solenoid. 
The CyDet must accurately and efficiently identify and measure signal electrons of 105\,MeV from aluminum target disks whilst suppressing backgrounds. 
It is designed to avoid high hit rates from the beam particles, background electrons from muon decay-in-orbit (DIO), and low energy protons emitted after the nuclear capture process of negative muons.

The momentum resolution of CDC should be less than 200\,keV/$c$ for the 105-MeV/$c$ electrons in order to distinguish the signal electrons from the DIO electrons. 
Note that the momentum resolution for the electrons at this momentum is dominated by the multiple scattering effect. 
In order to suppress this effect, \gasisobutane is selected as a gas mixture for its long radiation length of 1313\,m \cite{Moritsu:2018jdo}. 
He$-$iC$_{4}$H$_{10}$ was so far used to achieve high-momentum resolution in a low-momentum region in only a few experiments, e.g. KLOE \cite{Adinolfi:2002uk} and MEG-II \cite{Baldini:2018nnn} with a ratio of 90/10, and BaBar \cite{Aubert:2001tu} with 80/20.  

A major motivation of this work is to demonstrate that the current design of the cell and wire structure of the CDC with the \gasisobutane gas mixture can work in the COMET Phase-I experiment. 
We constructed a prototype chamber of the CDC and performed a beam test. 
In this paper, an experimental setup and an analysis method for the CDC prototype test are introduced. 
Then the drift space-time relation, drift velocity, spatial resolution, hit efficiency, gas gain and \dedx resolution are discussed.

\section{Experiment}

\subsection{A prototype chamber}

A detailed specification of CDC can be found in \cite{Adamov:2018vin}. 
The outer diameter, inner diameter and length of the CDC are 1.7~m, 1.0~m and 1.6~m, respectively, with a volume of 2~m$^3$. 
An alternating stereo wire configuration creates a total of 4986 drift cells with 20 concentric layers. 

\begin{figure}[htb!]
 \begin{center}
 \includegraphics[width=1.0\textwidth]{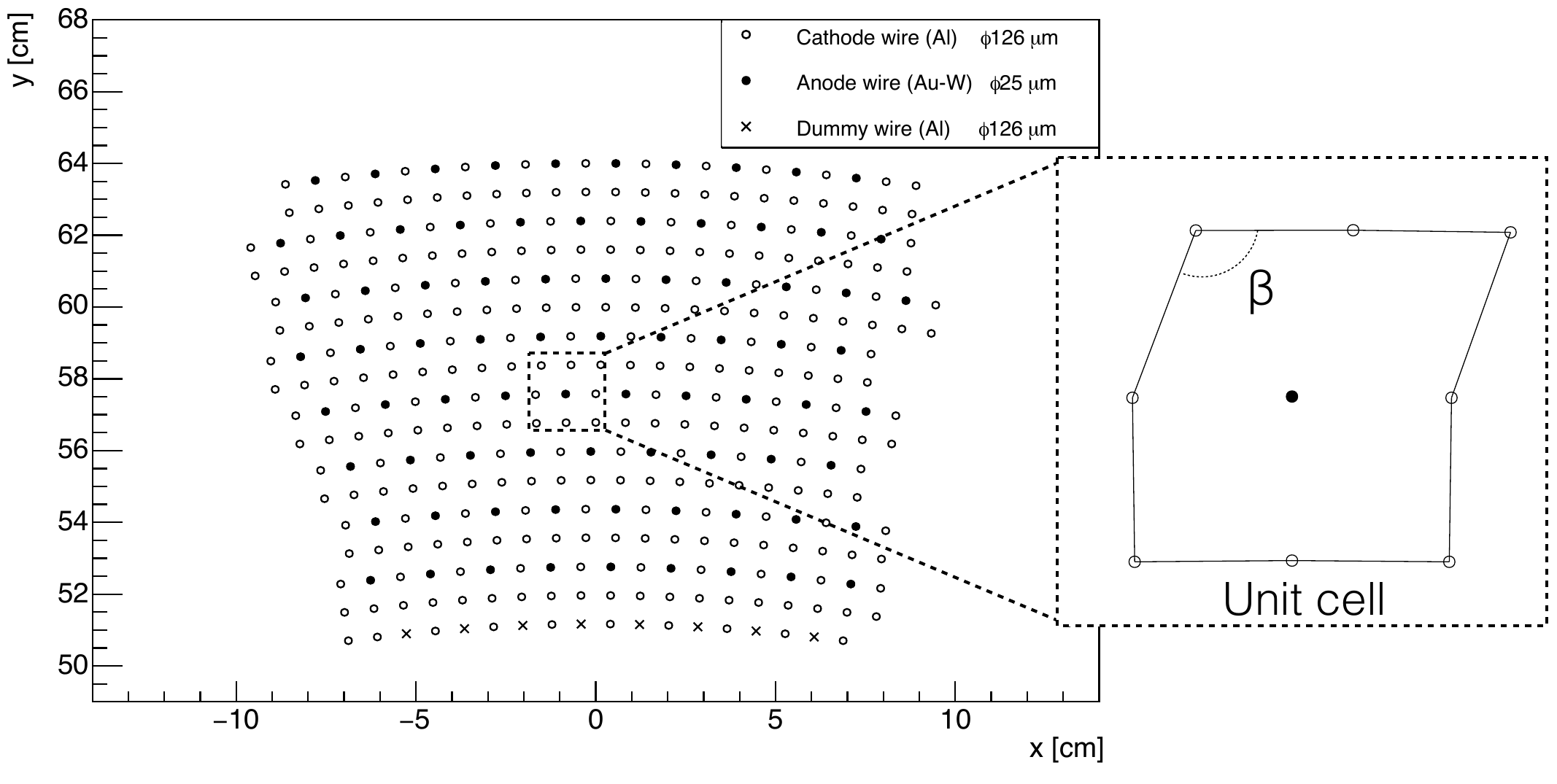}
 \end{center}
 \caption{Layout of wire configuration at the center of the chamber together with an expanded view of a unit cell. The cell deformation angle, $\beta$, ranges approximately 60 to 120 degrees depending on the $z$ position of the chamber.}
 \label{fig:layout_p4}
\end{figure}

\begin{figure}[htb!]
 \begin{center}
 \includegraphics[width=1.0\textwidth]{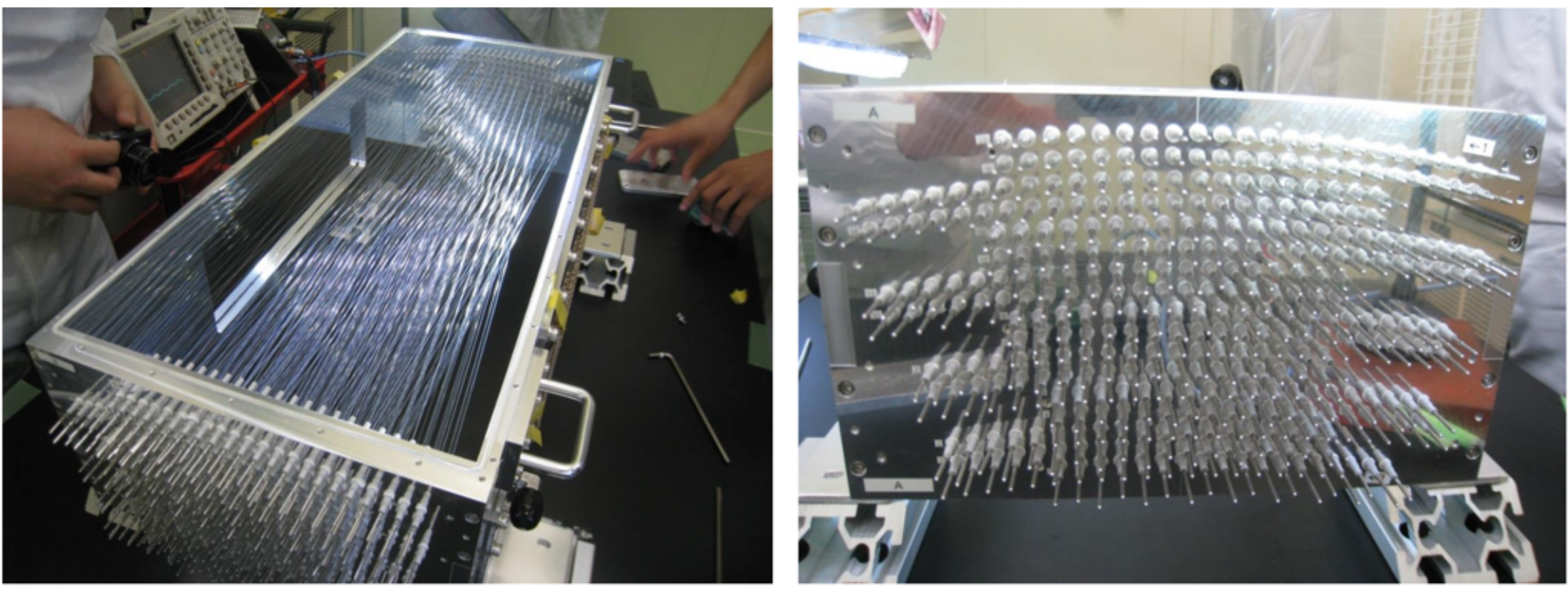}
 \end{center}
 \caption{Pictures of the prototype chamber with dimensions of $(600\times170\times280)$\,mm$^3$.}
 \label{fig:pic_p4}
\end{figure}

The prototype chamber was manufactured as a small partial copy of the CDC structure. 
It consists of 79 drift cells with 8 sensitive layers and 1 dummy layer. 
The purpose of sensitive layers is to detect the ionization electrons, whereas the dummy layer is used to study the effect of crosstalk from the electronics, which is not described in this paper. 
All the wires are strung with an alternating stereo angle of approximately $\pm 4$ degrees layer by layer. 
The cell shape is almost square with dimensions of (16.8$\times$16.0)\,mm$^2$. 
The layout of the cell structure at the center of the longitudinal direction is shown in Figure\,\ref{fig:layout_p4}, and pictures of the chamber are shown in Figure\,\ref{fig:pic_p4}. 
The size of the chamber is 600\,mm in length, 170\,mm in height, 280\,mm in width. 
There are 79 gold-plated tungsten wires with a diameter of 25\,$\mu$m as anode wires, and 265 aluminum wires with a diameter of 126\,$\mu$m as cathode wires. 

The endplates and walls of the chamber are made of aluminum. 
An engineering tolerance for the position of the wire ends with feedthroughs is less than 50\,$\mu$m. 
There are four aluminized Mylar windows with thickness of 200\,$\mu$m in all walls so that the effects of electromagnetic showers by the electron beam to the chamber can be minimized. 
A positive high voltage (HV) is supplied to the anode wires from one side of the endplates. 
From the other side, two front-end electronics boards are connected to the anode wires through decoupling capacitors. 
The electronics boards are covered with an aluminum box to suppress electromagnetic noise. 

The front-end readout electronics were fabricated based on the Readout Electronics for the Central Drift Chamber of Belle-II Experiment (RECBE) \cite{Uchida:2011ula,Taniguchi:2013pwa} with some modifications.
There are two outputs from the frontend amplifier--shaper--discriminator ASIC of RECBE \cite{Shimazaki:2014joa}: 
one is discriminated inside the ASIC and converted to timing information with 960-MHz TDC in an FPGA; 
therefore the time was digitized in a unit of 1.04~ns. 
The other is used to measure the charge with 10-bit, 30-MHz sampling rate ADC. 
Since the typical signal width is several tens of nanoseconds, the whole signal is sufficiently covered with the ADC time range of 1~$\mu$s. 
The SiTCP protocol \cite{Uchida:2008fha} is used to transmit the event data to a data-acquisition computer via Gigabit Ethernet fiber link.

\subsection{Experimental setup}

\begin{figure}[htb!]
 \begin{center}
 \includegraphics[width=0.8\textwidth]{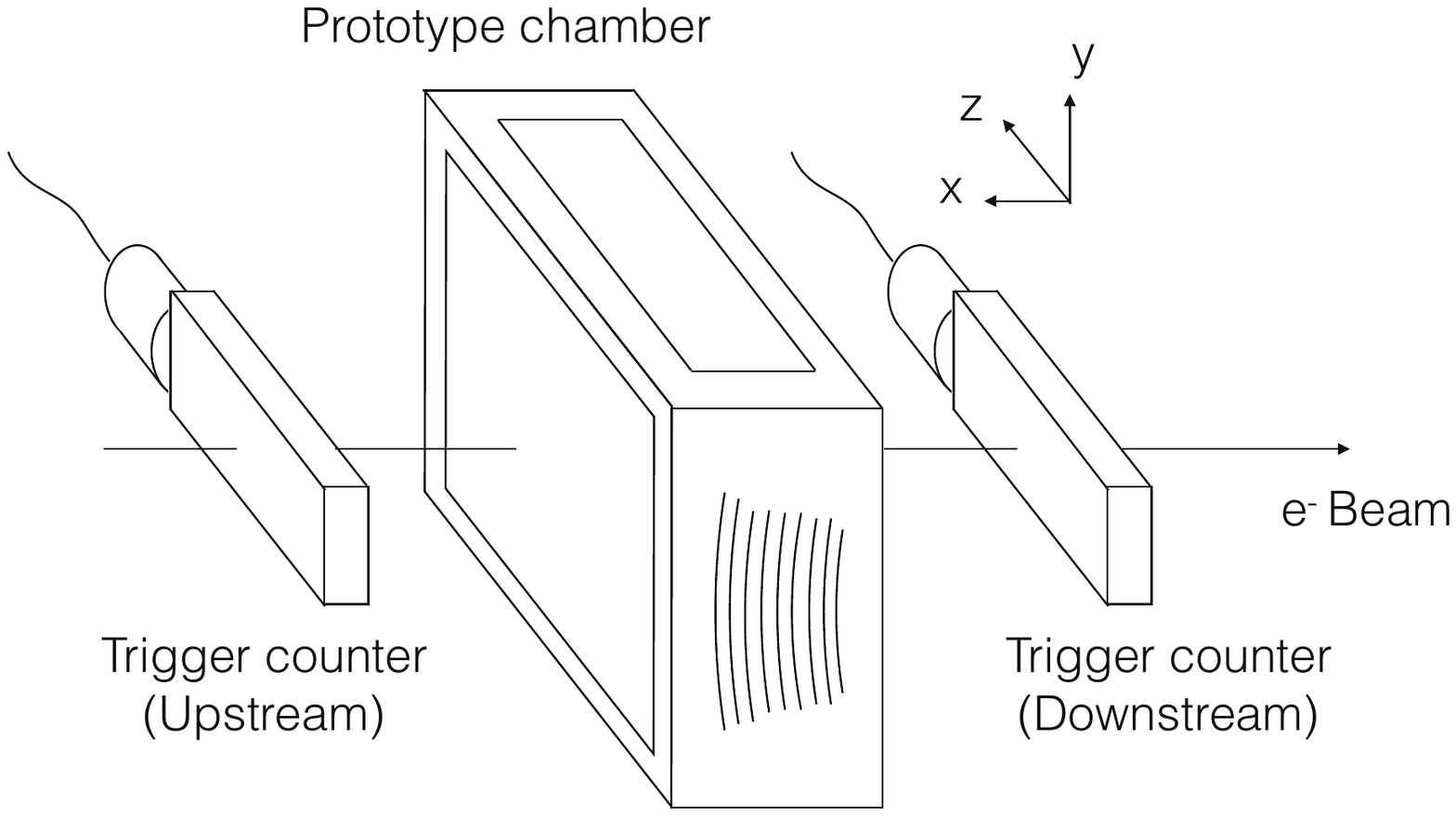}
 \includegraphics[width=0.8\textwidth]{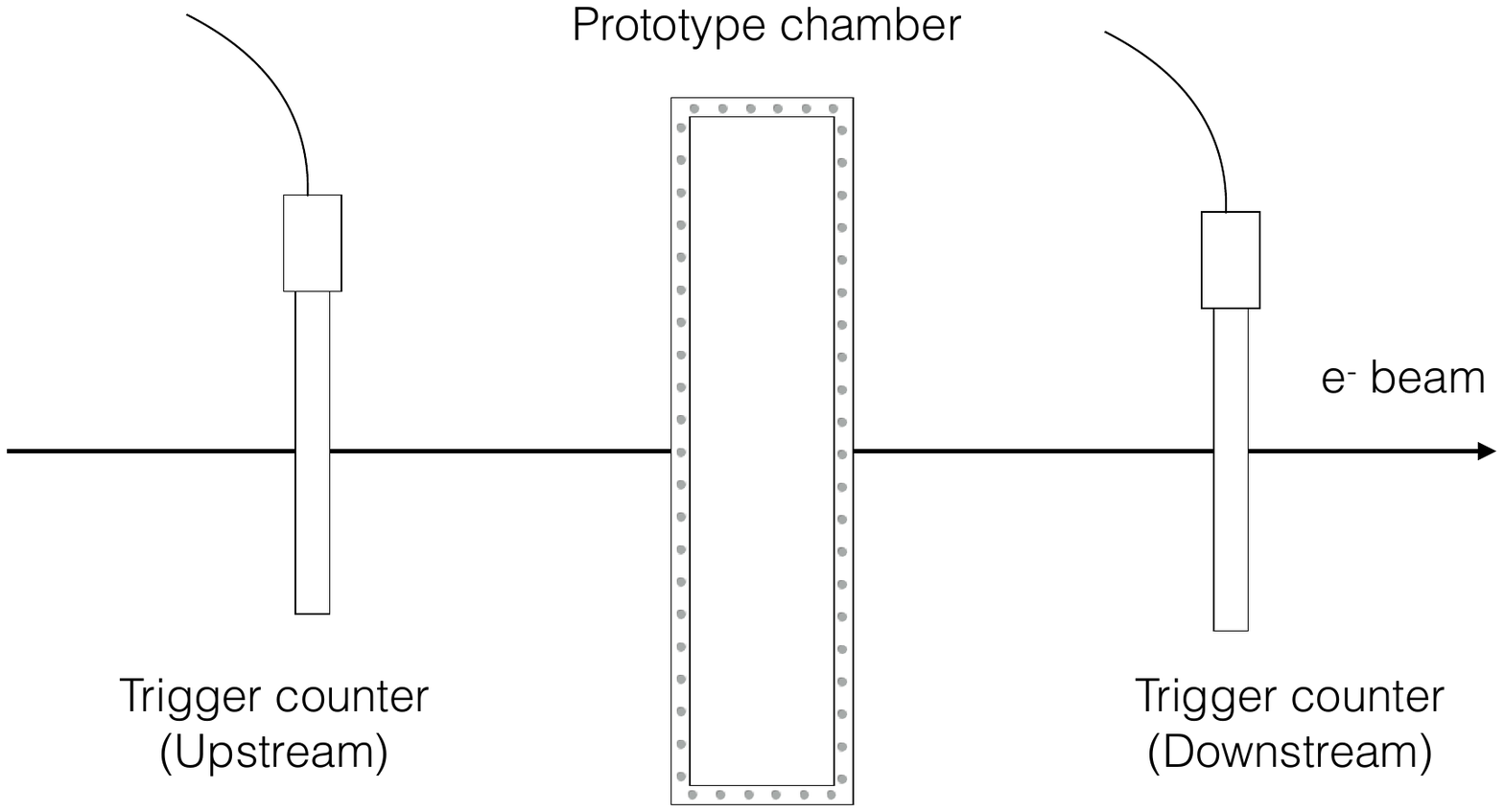}
 \end{center}
 \caption{\sl Schematic layouts of the experimental setup. The upper figure is a side view and the lower figure is a top view.}
 \label{fig:setup}
\end{figure}

The test experiment with an electron beam was performed in the Laser Electron Photon facility at SPring-8 (LEPS) \cite{Nakano:2001xp,muramatsu2012gev}, Japan. 
A photon beam of backward Compton scattering from UV laser, of 1.5 to 2.4\,GeV was used to generate a 1-GeV electron beam by colliding with a lead converter. 
The momentum selection of the electron beam was made with a dipole magnet located downstream of the lead converter. 
Two plastic scintillating counters of ($300\times90\times10$)\,mm$^{3}$ size were installed at both upstream and downstream of the prototype chamber. 
Schematic layouts of the experimental setup is shown in Figure\,\ref{fig:setup}. 
There was no significant magnetic field around the prototype chamber in this setup. 
The beam spreads over the whole chamber in the horizontal direction, whereas the vertical spread of the beam is around 5 to 6 cm. 

The timing and charge of wire hits were measured by RECBE and recorded by a data-acquisition computer using DAQ Middleware \cite{ref:DAQ_MW}. 
The event trigger was issued by a coincidence of signals from the two trigger counters. 
The average trigger rate was around 4--5 kHz. 
The high voltage and discrimination threshold were scanned at every 500 thousands events. 
A detailed list of the experimental conditions is shown in Table\,\ref{tb:CDC-prototype-run-sum}.

The gas flow rate is controlled at 20\,cc/min. 
As the gas gain of the chamber is sensitive to the environmental condition, the temperature of $26.0\pm0.5$\,$^{\circ}$C, the atmospheric pressure of $979\pm3$\,hPa, and the humidity of $53\pm5$\,\% were maintained during the tests. 
This level of variation in temperature and pressure affects the gas density at most by 0.5\%; 
and according to \cite{Blum:2008}, its effect on gas gain variation is estimated to be less than 4\%. 

\begin{table}[htbp]
 \centering
 \caption{\sl List of the experimental conditions. Here the discrimination threshold is defined as the analog output of the ASIC after amplification.}
 \label{tb:CDC-prototype-run-sum}
 \vspace{3mm}
 \begin{tabular}{lcc} \hline\hline
  Gas mixture & High voltage [V] & Threshold [mV] \\ \hline
  \gasisobutane & 1650--1830 & 10--30 \\
  \hline
 \end{tabular}
\end{table}

\section{Data Analysis}

\subsection{Track reconstruction}

\begin{figure}[htb!]
 \begin{center}
 \includegraphics[width=1.0\textwidth]{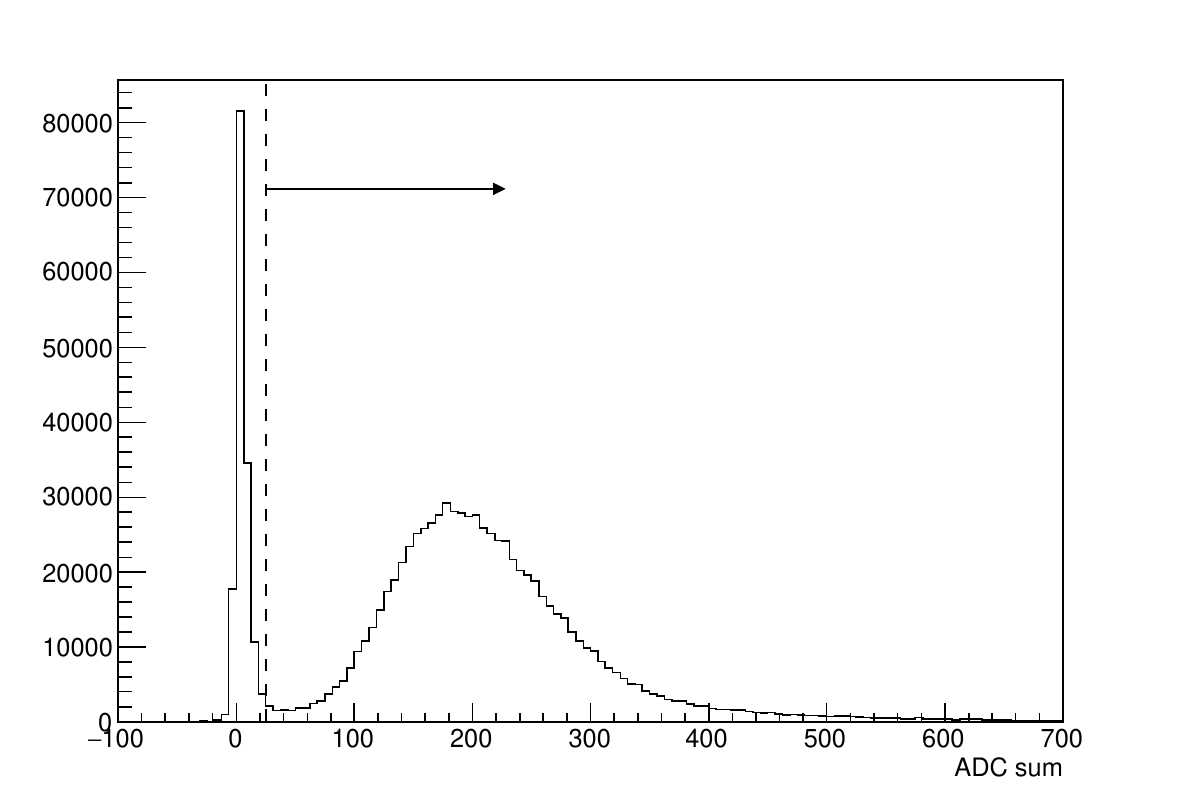}
 \end{center}
 \caption{\sl ADC sum distribution measured at a drift cell for \gasisobutane at 1800\,V with 10-mV threshold. In this case the ADC sum larger than 25 was selected, as indicated by the dashed line.}
 \label{fig:adc}
\end{figure}

The signal waveforms were recorded in 32 sampling points of ADC with a 33-ns interval. 
The ADC values of each channel were numerically summed up separately after subtracting a pedestal level to obtain charge information; 
we call this as ADC sum hereafter. 
In the beginning of the data analysis, the wire hits were classified by their ADC sum. 
A typical ADC sum distribution of a cell is shown in \Cref{fig:adc}. 
Since signals and noises were clearly separated, a cut for signal hit identification was applied based on the ADC sum for each drift cell. 
We used a calibration function obtained in an earlier electronics test for converting the ADC sum to charge. 

In order to test the performance of a specific cell and layer, the hits in the dummy layer and a layer to be tested were excluded from the track reconstruction to avoid bias. 
Events with the total number of signal hits larger than 6 were used for track reconstruction. 
Using a drift space-time relation, which will be discussed in the following subsection, drift time was converted into the drift distance, $d$. 
The residual of a signal hit was defined as $r = d - d^\textrm{fit}$, where $d^\textrm{fit}$ is the distance of the closest approach from the reconstructed track to the anode wire in the hit cell. 
Excluding the layer to be tested, the sum of normalized residuals of signal hits was defined as the $\chi^{2}$ of the reconstructed track: 
\begin{equation}
  \chi^{2} = \sum_{i} \Bigg(\frac{d_i - d_i^\textrm{fit}} {\sigma(d_i)}\Bigg)^2 , 
  \label{eqn:min}
\end{equation}
where $i$ is a hit index, and $\sigma(d)$ denotes spatial resolution as a function of the distance $d$, the details of which will be discussed in Section~\ref{sec:spatialreso}. 
A $\chi^{2}$--minimization algorithm was applied to obtain a realistic reconstructed track. 
In the following study, the central forth layer was chosen as the layer to be tested, so as to minimize a tracking error, unless otherwise noted.

\subsection{Drift space-time relation}

\begin{figure}[h]
 \begin{center}
 \includegraphics[width=0.8\textwidth]{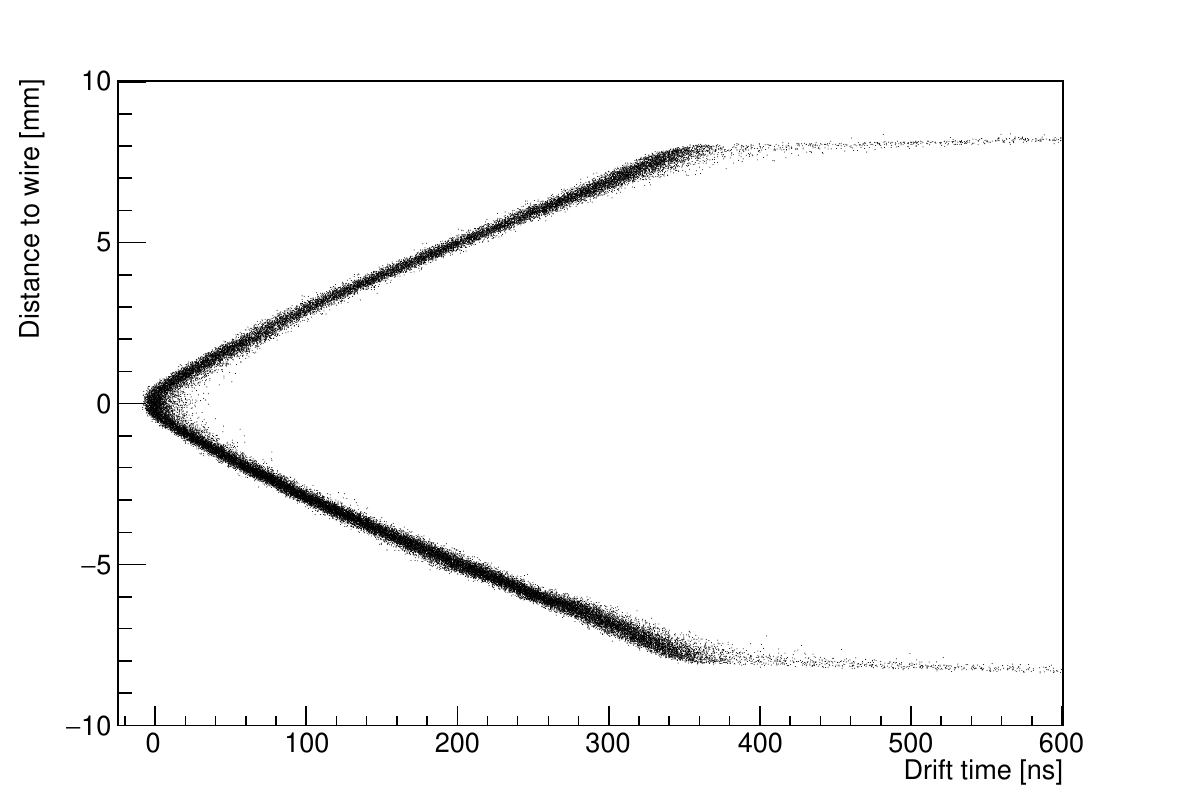}
 \end{center}
 \caption{\sl Space-time distribution of the central layer obtained for \gasisobutane at 1800\,V with 10\,mV threshold.}
 \label{fig:xtd}
\end{figure}

\begin{figure}[h]
 \begin{center}
 \includegraphics[width=1.0\textwidth]{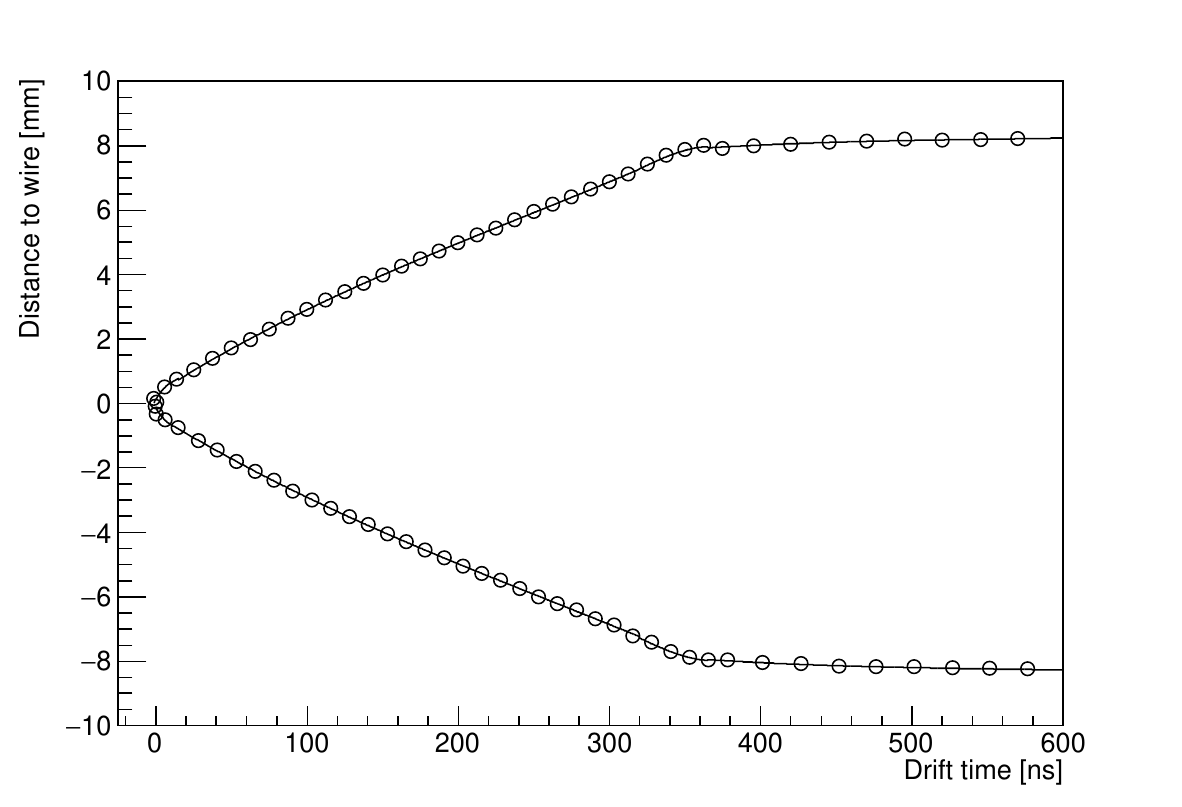}
 \end{center}
 \caption{\sl Fitted space-time relation for \gasisobutane at 1800\,V with 10-mV threshold. 
The open circles represent the averaged space-time distribution, and the solid line is the fitted function. 
Notice that the number of circles shown here is less than the used intervals for a better visualization.}
 \label{fig:xtfitted}
\end{figure}

A drift space-time (hereafter \strel) relation was studied by an iterative procedure starting from an initial value obtained from Garfield++ \cite{ref:gpp}. 
A typical \strel distribution is shown in \Cref{fig:xtd}. 
The \strel relation was calibrated layer by layer. 
The distribution was averaged over 200-$\mu$m intervals near the anode wire and over 3-ns intervals near the cell boundary. 
To get a parameterized \strel relation, the averaged distribution was then fitted by piecewise continuous third-order polynomial functions connecting four different drift time regions. 
An example of the \strel relation fitting is shown in \Cref{fig:xtfitted}. 
The \strel relations for data sets with different high voltages or thresholds were calibrated separately, and applied for the following analysis procedures accordingly. 

\subsection{Drift velocity}

\begin{figure}[h]
 \begin{center}
 \includegraphics[width=1.0\textwidth]{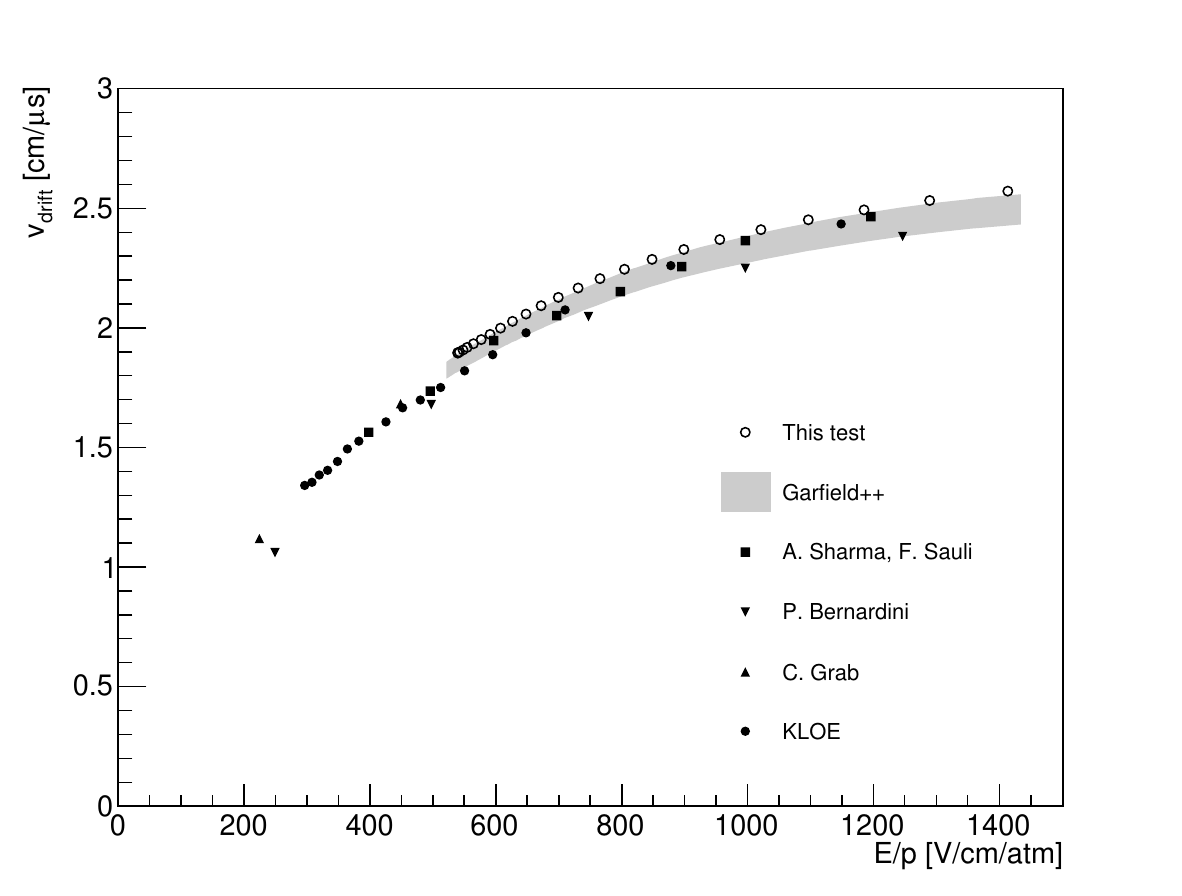}
 \end{center}
 \caption{\sl Drift velocity versus the electric field for \gasisobutane at 1~atm by comparing with Garfield++ simulation, calculations of Sharma--Sauli~\cite{Sharma:1994rc}, and experiments of P.~Bernardini~\cite{Bernardini:1994db}, C.~Grab~\cite{Grab:1992ej}, and a KLOE-CDC prototype~\cite{Avanzini:2000ve}. 
For the band of Garfield++ simulation, the lower boundary corresponds to the gas mixture ratio between He and iC$_{4}$H$_{10}$ at $90.4:9.6$, and the upper boundary corresponds to the gas mixture ratio at $89.6:10.4$ with 0.12\% water vapor.}
 \label{fig:dvde}
\end{figure}

The drift velocity was calculated by taking derivatives of the \strel relation as a function of the distance from the wire. 
The local electric field in a drift cell was calculated by using the Garfield++ simulation. 
Then we obtained the drift velocity as a function of the local electric field as shown in \Cref{fig:dvde}.
The drift velocity in \gasisobutane gas mixture is unsaturated near 1000~V/cm/atm. 
\Cref{fig:dvde} also shows other results of calculations~\cite{Sharma:1994rc} and experimental measurements~\cite{Bernardini:1994db,Grab:1992ej,Avanzini:2000ve} together with Garfield++ simulation.
In the Garfield++ simulation, we considered an experimental uncertainty of gas mixture ratio as ranging from $89.6:10.4$ to $90.4:9.6$, and an uncertainty of water vapor contamination as ranging from 0 to 0.12\%~
\footnote{The experimental uncertainty of the gas mixture ratio is given by specifications of mass flow controllers; and the uncertainty of the water vapor contamination is estimated by some tests under different circumstances after the beam test.}.
All data show good agreement; the minimal deviations could be justified within small differences in operating conditions, such as pressure, temperature, gas mixture ratio, and water vapor contamination. 

\section{Results and Discussion}

\subsection{Scan of applied high voltage}

In order to determine an optimal operation high voltage, the spatial resolution and cell hit efficiency were studied as a function of applied high voltage. 
The intrinsic spatial resolution was extracted from the standard deviation of residual distributions, $\sigma_{\textrm{total}}$, by subtracting a tracking error in quadrature. 
Since the tracking error is caused by extrapolating or interpolating the reconstructed track, it was estimated from a simulation with assumed intrinsic spatial resolution. 
Comparing the simulation with the experimental data, the tracking error was derived to be 94\,$\mu$m for the central layer. 
The cell hit efficiency was defined as the ratio of the number of hits with the residual within $\pm 5\, \sigma_{\textrm{total}}$ to the number of tracks passing through the cell. 

\begin{figure}[htb!]
 \begin{center}
   \includegraphics[width=1.0\textwidth]{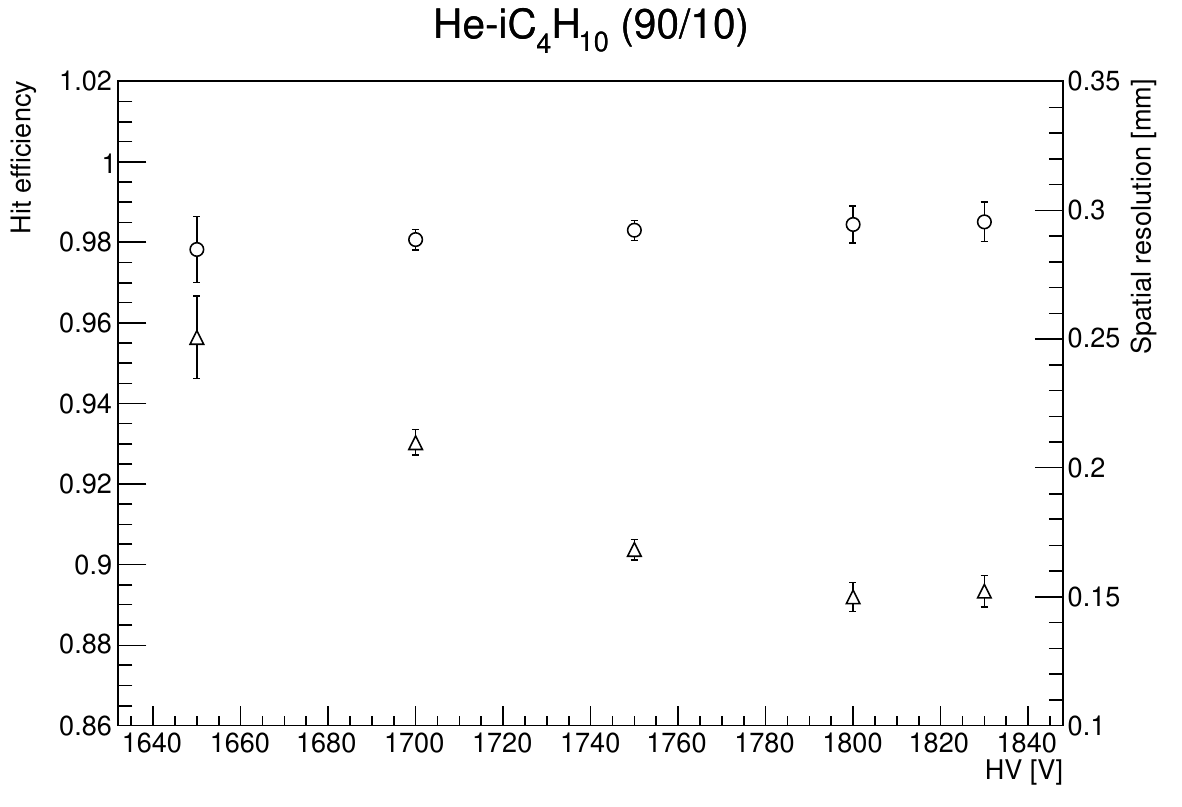}
 \end{center}
 \caption{\sl Cell hit efficiency (circles) and intrinsic spatial resolution (triangles) as a function of the applied voltage for \gasisobutane.}
 \label{fig:eff_res_hv}
\end{figure}

The results are shown in \Cref{fig:eff_res_hv} for high voltages from 1650 to 1830~V. 
Note that an optimal threshold value which gave the best spatial resolution was selected for each high-voltage data set. 
The plateau-like behavior of efficiency and spatial resolution are consistent with other literatures \cite{Adinolfi:2002uk,Uno:1992rs}. 
We obtained the best spatial resolution of 150\,$\mu$m with 99\% hit efficiency at the voltages higher than 1800\,V. 

\subsection{Spatial resolution}
\label{sec:spatialreso}

The spatial resolution varies depending on the distance from the anode wire. 
Detailed contributions to spatial resolution were studied using a similar method of \cite{Avanzini:2000ve}. 
The components of spatial resolution can be grouped into primary ionization statistics ($\sigma_{\textrm{ion}}$), electron longitudinal diffusion ($\sigma_{\textrm{diff}}$), and electronics time resolution ($\sigma_{\textrm{ele}}$): 
\begin{equation}
  \sigma = \sqrt{\sigma_{\textrm{ion}}^2 + \sigma_{\textrm{diff}}^2 + \sigma_{\textrm{ele}}^2}. 
  \label{eqn:totalres}
\end{equation}

The contribution of primary ionization statistics was estimated from a Monte Carlo simulation. 
Primary ions were generated according to $\exp(-n_\textrm{p} y)$, where $n_\textrm{p}$ is the number of primary ions per unit length, and $y$ is the track path length in a drift cell. 
$\sigma_{\textrm{ion}}$ was estimated from the standard deviation of the difference between the distance from anode wire to the track and to the closest ionization point. 
Therefore, $\sigma_{\textrm{ion}}$ was obtained as a function of the distance to wire and $n_\textrm{p}$. 

The function of $\sigma_{\textrm{diff}}$ follows an approximation of a classical diffusion theory based on the Boltzmann statistics: $\sigma_{\textrm{diff}}(x) = \sqrt{(2Dx/v)}$, where $D$, $x$, and $v$ are the diffusion coefficient, drift distance, and average drift velocity, respectively. 
For $\sigma_{\textrm{ele}}$, the main contributions come from the time resolution of the readout electronics and time walk effect of signals. 
The time resolution was determined from specification of the electronics to be $1.04/\sqrt{12}$~ns, which corresponds to 10\,$\mu$m, whereas the time walk effect was found to be negligible by choosing a low threshold level of electronics. 

\begin{figure}[htb!]
 \begin{center}
   \includegraphics[width=1.0\textwidth]{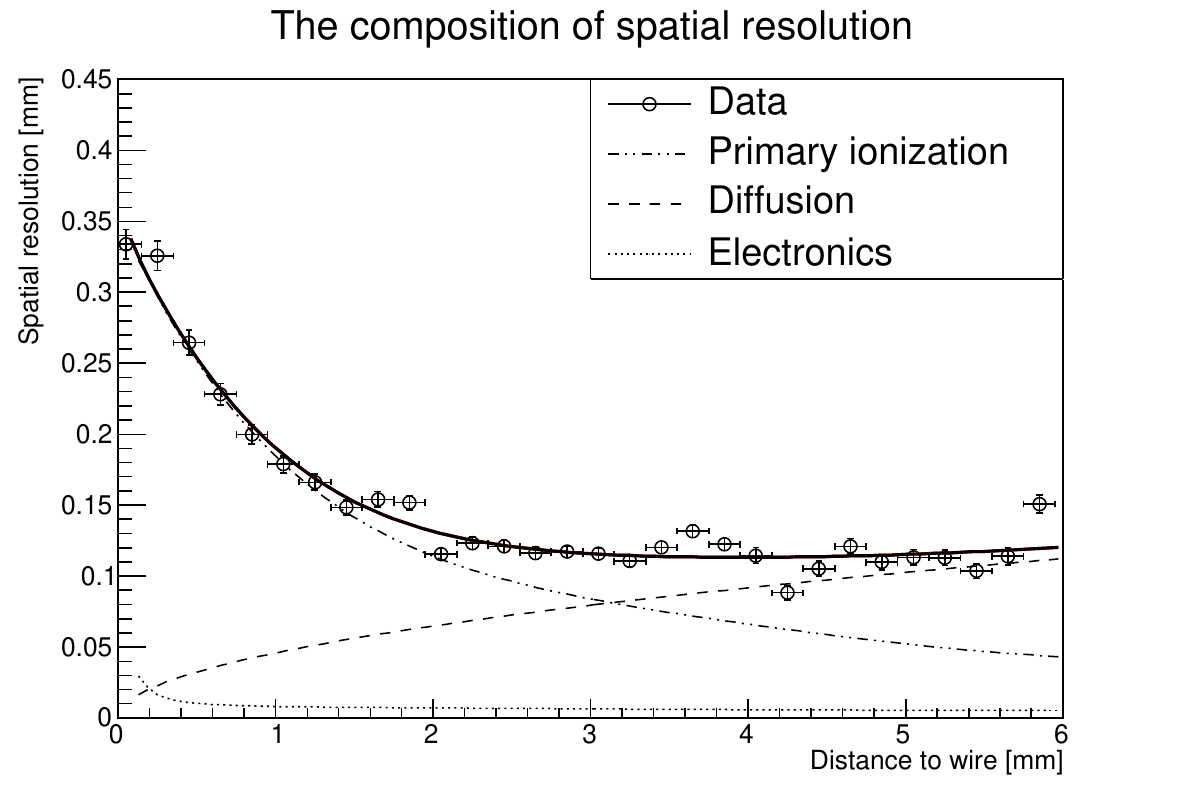}
 \end{center}
 \caption{\sl Spatial resolution as a function of distance to the anode wire for \gasisobutane at 1800\,V with 10-mV threshold. 
The open circles represent the experimental data. The solid line indicates the fitting results. 
The dot--dashed line, dashed line and dotted line show the contribution of the primary ionization, longitudinal diffusion, and electronics time resolution, respectively.}
 \label{fig:fitting_res}
\end{figure}

\Cref{fig:fitting_res} shows the spatial resolution as a function of distance to the anode wire together with the fitting result by Eq.\ref{eqn:totalres}. 
The fitting result shows that the number of primary ionization $n_\textrm{p} = 14.2 \pm 0.2$ cm$^{-1}$. 
For comparison with other results, the $n_\textrm{p}$ value was scaled from the 1-GeV electron in our experiment to minimum ionizing particles taking the energy loss difference into account. 
Using the photo-absorption ionization model \cite{Apostolakis:2000yu}, we obtained $n_\textrm{p} = 9.9 \pm 0.1$\,cm$^{-1}$ for minimum ionizing particles. 
Note that the uncertainty is only statistical, and systematic uncertainties are not taken into account. 
The result is slightly smaller but in a decent agreement with 12.7\,cm$^{-1}$ computed by Sharma and Sauli~\cite{Sharma:1994rc} and $12.3 \pm 0.2$\,cm$^{-1}$ reported in ~\cite{Avanzini:2000ve}. 

The fitting result also gives $D/v = (1.05 \pm 0.04) \times 10^{-4}$ cm for the ratio of the diffusion coefficient to the drift velocity. 
The value corresponds to 112\,$\mu$m for 0.6\,cm of drift. 
It was found that the deterioration of the spatial resolution near the anode wire is due to the statistical fluctuation of primary ionization, whereas the longitudinal diffusion dominates in the long drift distance.

\subsection{Gas gain}

\begin{figure}[h]
 \begin{center}
 \includegraphics[width=1.0\textwidth]{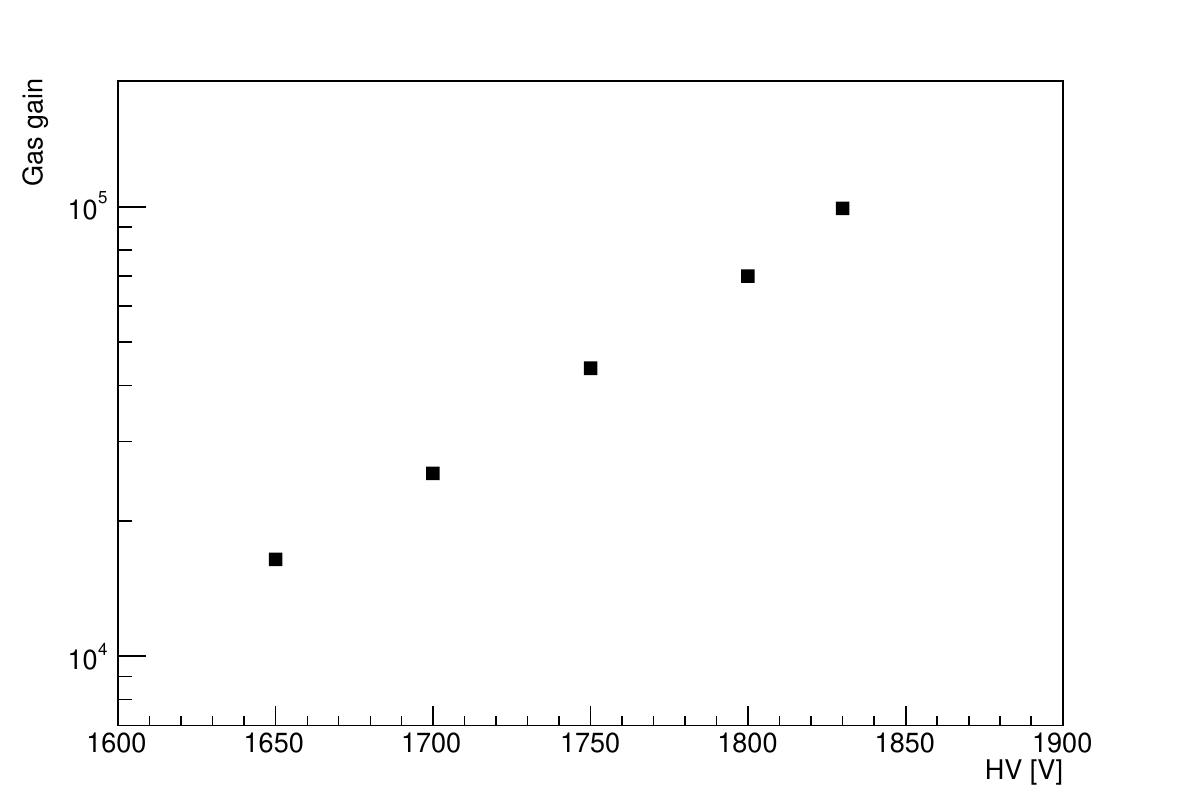}
 \end{center}
 \caption{\sl Relation between gas gain and high voltage for \gasisobutane with 10-mV threshold.}
 \label{fig:gasgainHVScan}
\end{figure}

Once the track was reconstructed,
the measured charge at each drift cell was then divided by the path length of the track in the drift cell (hereafter $Q$).
The gas gain $G$ is computed as $Q/(n_\textrm{t} e)$, where $n_\textrm{t}$ is the total number of electron--ion pairs per unit length, and $e$ denotes the elementary charge.
To avoid the non-linear effect in the Landau tail of the distribution of $Q$ and $n_\textrm{t}$ contributed by the electronics and the space charge effect, we took the most probable values (MPV) for each distribution in one drift cell.
The $n_\textrm{t}^\textrm{MPV}$ was computed as $(dE/dx)^\textrm{MPV}/W_\textrm{I}$, where $(dE/dx)^\textrm{MPV}$ is the most probable energy deposition per unit length in one drift cell, and $W_\textrm{I}$ is the average ionization energy per ion pair.
Ignoring the Penning effect, $W_\textrm{I}$ in \gasisobutane was computed as 29.3\,eV based on the He and iC$_{4}$H$_{10}$ properties given in \cite{Tanabashi:2018oca}.
According to the photon absorption ionization model \cite{Apostolakis:2000yu}, $(dE/dx)^\textrm{MPV}$ for an 1-GeV electron in this gas mixture is computed to be 0.70\,keV/cm, therefore $n_\textrm{t}^\textrm{MPV}$ = 23.9\,cm$^{-1}$.
The gas gain was then calculated for data sets at different high voltages using the measured $Q^\textrm{MPV}$.
\Cref{fig:gasgainHVScan} shows the relationship between the gas gain and the applied high voltage. 
A linear relation in a logarithmic scale can be observed as expected. 

\subsection{\dedx resolution}

\begin{figure}[h]
 \begin{center}
 \includegraphics[width=1.0\textwidth]{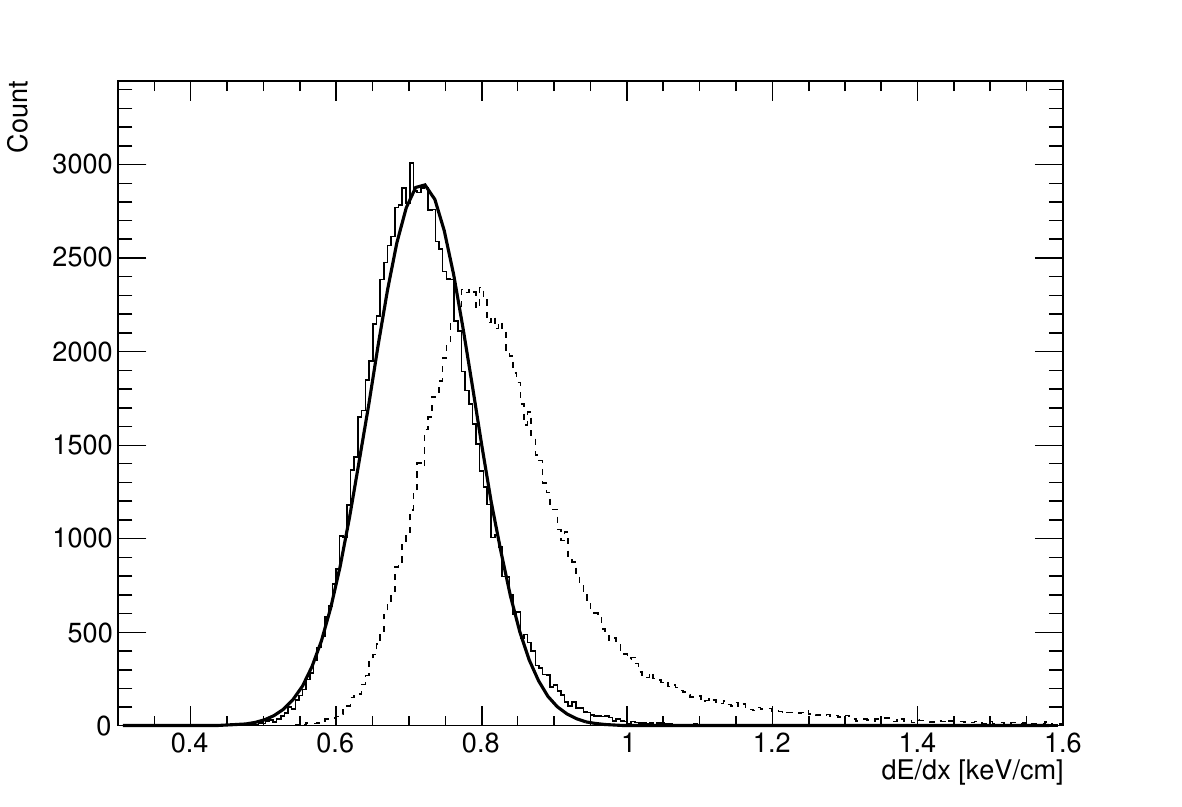}
 \end{center}
 \caption{\sl \dedx distribution measured for tracks with 10 samples using data taken from \gasisobutane at 1800\,V with 10-mV threshold. The solid and dashed lines denote histograms with 80$\,\%$ truncated mean and without truncation, respectively.}
 \label{fig:dedx100VS90}
\end{figure}

\begin{figure}[h]
 \begin{center}
 \includegraphics[width=1.0\textwidth]{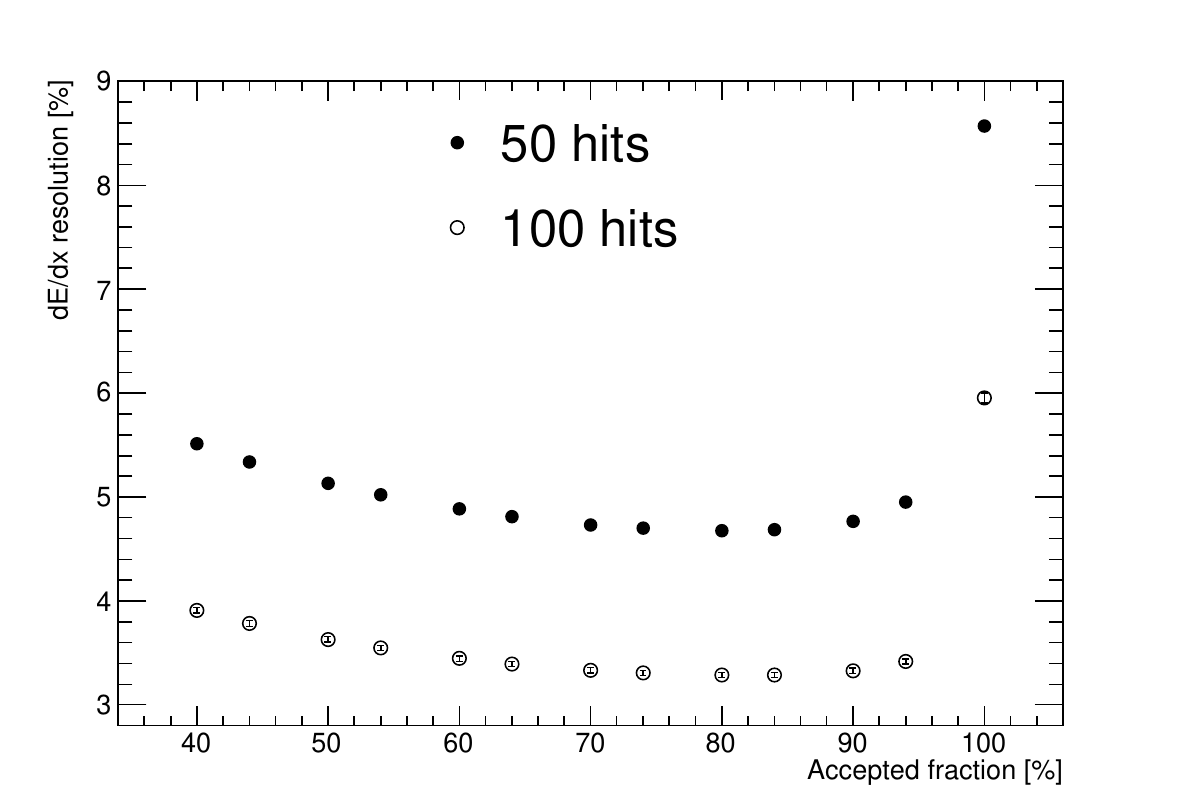}
 \end{center}
 \caption{\sl \dedx resolution for tracks with 50 and 100 samples using data taken from \gasisobutane at 1800\,V with 10-mV threshold as a function of the truncation ratio.}
 \label{fig:dedxResVSfrac}
\end{figure}

For the measurement of the \dedx resolution,
tracks with arbitrary lengths by combining hits in reconstructed events were simulated.
To avoid Landau tail of collected charge distribution worsening the \dedx resolution, the truncated mean technique was applied \cite{Andryakov:1998yk}.
A comparison of the \dedx distribution with and without truncation is shown in \Cref{fig:dedx100VS90}.
We found that the best \dedx resolution can be achieved by accepting 80$\,\%$ of hits in a track, as shown in \Cref{fig:dedxResVSfrac}.

\begin{figure}[h]
 \begin{center}
 \includegraphics[width=1.0\textwidth]{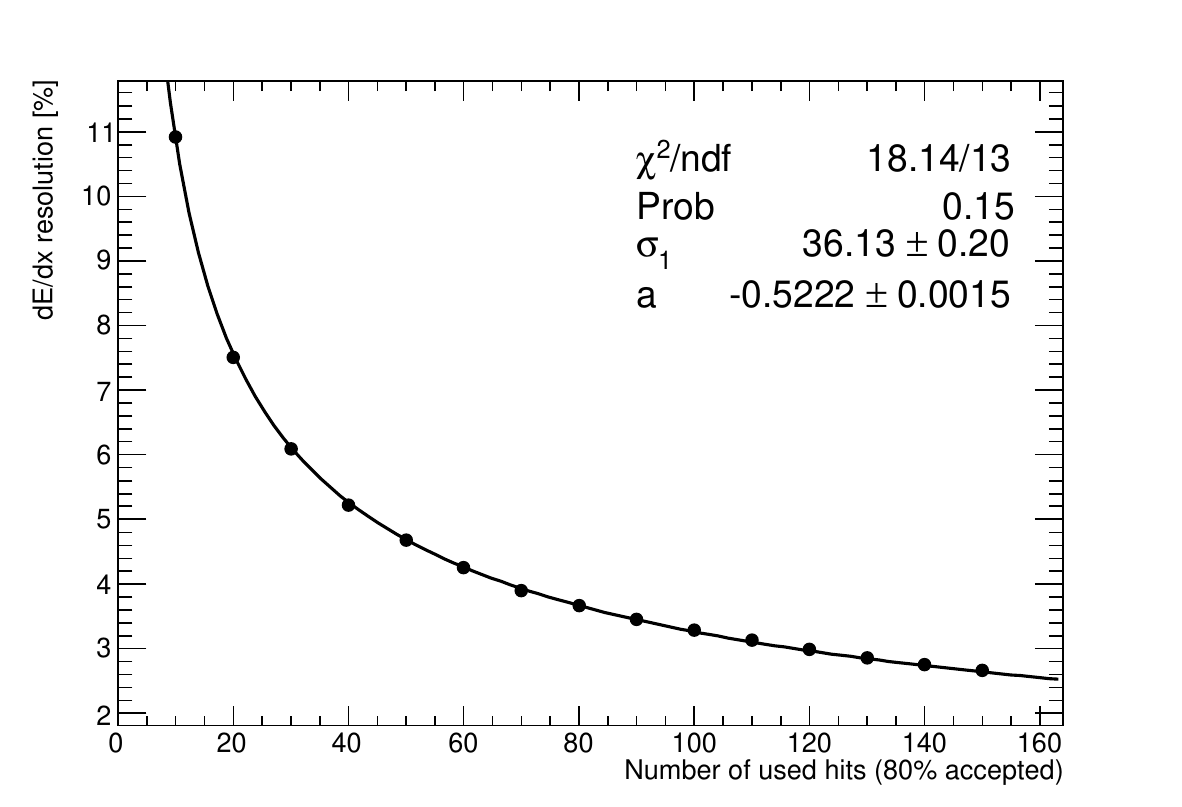}
 \end{center}
 \caption{\sl \dedx resolution of tracks with different number of samples taken from \gasisobutane at 1800\,V with 10\,mV threshold.  The solid line denotes the fitting result with Eq.~\ref{eqn:dedxreso}; here $\sigma_0 x^b$ is absorbed into $\sigma_1$ with $x=1.6$\,cm.}
 \label{fig:dedxResVSN}
\end{figure}

The \dedx resolution can be parameterized as \cite{Allison:1980vw} 
\begin{equation}
  \sigma_{\textrm{d}E/\textrm{d}x} = \sigma_{0}N^{a}x^{b} , 
  \label{eqn:dedxreso}
\end{equation}
where $N$ is the number of samples and $x$ is the track length.
By varying the number of samples included in the combined tracks, a relationship between \dedx resolution and $N$ was obtained  as shown in \Cref{fig:dedxResVSN}. 
Since a typical number of hits in the CDC for a \mue conversion signal track is around 50, the predicted \dedx resolution in COMET Phase-I was found to be $4.7\,\%$ for 1-GeV electrons with an assumption of track length per cell $x = 1.6$\,cm.

\section{Conclusions and Outlook}

We have constructed a prototype chamber for the CDC used in the COMET Phase-I experiment. 
The cell and wire structure of the prototype is identical with the CDC. 
A test experiment was carried out using a 1-GeV electron beam at SPring-8. 
Experimental data were obtained with a gas mixture of \gasisobutane for a couple of high voltages and discrimination thresholds of electronics without a magnetic field. 
We confirmed a similar drift velocity with other literatures. 
The spatial resolution of 150\,$\mu$m with the hit efficiency of $99\,\%$ was obtained at applied voltages higher than 1800~V. 
The behavior of spatial resolution with respect to the distance from the anode wire was understood adequately. 
The gas gain behavior with respect to the high voltages was also observed as expected. 
The \dedx resolution was estimated to be 4.7\% assuming a typical \mue conversion signal in COMET Phase-I. 

We have demonstrated that the CDC design with the \gasisobutane gas mixture provides sufficient performance in the COMET Phase-I experiment. 
It should be noted that the CDC will be operated in a 1-T magnetic field and the track injection angles are not always orthogonal to the cell, which makes different drift space-time relation from this test. 
However, these effects are expected to be controlled by carefully adjusting the space-time relation. 
Detailed simulation study for the real operating condition of the CDC can be found in \cite{Adamov:2018vin}. 

Based on the test results, the construction of the CDC has been successfully completed, and the performance tests using cosmic rays are in progress. 
Preliminary reports are available in \cite{Moritsu:2018jdo,Moritsu:2019uoy}, and publication of the full report is in preparation.

\section*{Acknowledgements}

The experiment was performed at the BL33LEP of SPring-8 with the approval of the Japan Synchrotron Radiation Research Institute (JASRI)  as a contract beamline (Proposal No.~BL33LEP/6001-Q038). 
We are very thankful for staffs and students in the LEPS facility for their support during the beam time. 
This work of the Osaka University group was supported in part by the Japan Society of the Promotion of Science (JSPS) KAKENHI Grant No.~25000004 and 18H05231; and the Chinese Academy of Sciences (CAS) President’s International Fellowship Initiative under Contract No.~2018PM0004. 
The work of the Institute of High Energy Physics (IHEP) was support in part by the National Natural Science Foundation of China (NSFC) under Contracts No.~11335009 and 11475208; Research Program of IHEP under Contract No.~Y3545111U2; and the State Key Laboratory of Particle Detection and Electronics of IHEP, China, under Contract No.~H929420BTD.


\bibliography{mybibfile}

\begin{thebibliography}{10}
\expandafter\ifx\csname url\endcsname\relax
  \def\url#1{\texttt{#1}}\fi
\expandafter\ifx\csname urlprefix\endcsname\relax\def\urlprefix{URL }\fi
\expandafter\ifx\csname href\endcsname\relax
  \def\href#1#2{#2} \def\path#1{#1}\fi

\bibitem{Bertl:2006up}
W.~H. Bertl, et~al., {A Search for muon to electron conversion in muonic gold},
  Eur. Phys. J. C 47 (2006) 337--346.
\newblock \href {http://dx.doi.org/10.1140/epjc/s2006-02582-x}
  {\path{doi:10.1140/epjc/s2006-02582-x}}.

\bibitem{Kuno:2013mha}
Y.~Kuno, {A search for muon-to-electron conversion at J-PARC: The COMET
  experiment}, PTEP 2013 (2013) 022C01.
\newblock \href {http://dx.doi.org/10.1093/ptep/pts089}
  {\path{doi:10.1093/ptep/pts089}}.

\bibitem{Adamov:2018vin}
R.~Abramishvili, et~al., {COMET Phase-I Technical Design Report}, PTEP 2020~(3)
  (2020) 033C01.
\newblock \href {http://arxiv.org/abs/1812.09018} {\path{arXiv:1812.09018}},
  \href {http://dx.doi.org/10.1093/ptep/ptz125}
  {\path{doi:10.1093/ptep/ptz125}}.

\bibitem{Moritsu:2018jdo}
M.~Moritsu, et~al., {Construction and performance tests of the COMET CDC}, PoS
  ICHEP2018 (2019) 538.
\newblock \href {http://dx.doi.org/10.22323/1.340.0538}
  {\path{doi:10.22323/1.340.0538}}.

\bibitem{Adinolfi:2002uk}
M.~Adinolfi, et~al., {The tracking detector of the KLOE experiment}, Nucl.
  Instrum. Meth. A 488 (2002) 51--73.
\newblock \href {http://dx.doi.org/10.1016/S0168-9002(02)00514-4}
  {\path{doi:10.1016/S0168-9002(02)00514-4}}.

\bibitem{Baldini:2018nnn}
A.~M. Baldini, et~al., {The design of the MEG II experiment}, Eur. Phys. J. C
  78~(5) (2018) 380.
\newblock \href {http://arxiv.org/abs/1801.04688} {\path{arXiv:1801.04688}},
  \href {http://dx.doi.org/10.1140/epjc/s10052-018-5845-6}
  {\path{doi:10.1140/epjc/s10052-018-5845-6}}.

\bibitem{Aubert:2001tu}
B.~Aubert, et~al., {The BaBar detector}, Nucl. Instrum. Meth. A 479 (2002)
  1--116.
\newblock \href {http://arxiv.org/abs/hep-ex/0105044}
  {\path{arXiv:hep-ex/0105044}}, \href
  {http://dx.doi.org/10.1016/S0168-9002(01)02012-5}
  {\path{doi:10.1016/S0168-9002(01)02012-5}}.

\bibitem{Uchida:2011ula}
T.~Uchida, et~al., {Readout electronics for the central drift chamber of the
  Belle II detector}, in: {2011 IEEE Nuclear Science Symposium and Medical
  Imaging Conference}, 2011, pp. 694--698.
\newblock \href {http://dx.doi.org/10.1109/NSSMIC.2011.6154084}
  {\path{doi:10.1109/NSSMIC.2011.6154084}}.

\bibitem{Taniguchi:2013pwa}
N.~Taniguchi, et~al., {All-in-one readout electronics for the Belle-II Central
  Drift Chamber}, Nucl. Instrum. Meth. A 732 (2013) 540--542.
\newblock \href {http://dx.doi.org/10.1016/j.nima.2013.06.096}
  {\path{doi:10.1016/j.nima.2013.06.096}}.

\bibitem{Shimazaki:2014joa}
S.~Shimazaki, T.~Taniguchi, T.~Uchida, M.~Ikeno, N.~Taniguchi, M.~M. Tanaka,
  {Front-end electronics of the Belle II drift chamber}, Nucl. Instrum. Meth. A
  735 (2014) 193--197.
\newblock \href {http://dx.doi.org/10.1016/j.nima.2013.09.050}
  {\path{doi:10.1016/j.nima.2013.09.050}}.

\bibitem{Uchida:2008fha}
T.~Uchida, {Hardware-Based TCP Processor for Gigabit Ethernet}, IEEE Trans.
  Nucl. Sci. 55~(3) (2008) 1631--1637.
\newblock \href {http://dx.doi.org/10.1109/TNS.2008.920264}
  {\path{doi:10.1109/TNS.2008.920264}}.

\bibitem{Nakano:2001xp}
T.~Nakano, et~al., {Multi-GeV laser electron photon project at SPring-8}, Nucl.
  Phys. A 684 (2001) 71--79.
\newblock \href {http://dx.doi.org/10.1016/S0375-9474(01)00490-0}
  {\path{doi:10.1016/S0375-9474(01)00490-0}}.

\bibitem{muramatsu2012gev}
N.~Muramatsu, Gev photon beams for nuclear/particle physics (2012).
\newblock \href {http://arxiv.org/abs/1201.4094} {\path{arXiv:1201.4094}}.

\bibitem{ref:DAQ_MW}
Y.~{Yasu}, K.~{Nakayoshi}, E.~{Inoue}, H.~{Sendai}, H.~{Fujii}, N.~{Ando},
  T.~{Kotoku}, S.~{Hirano}, T.~{Kubota}, T.~{Ohkawa}, A data acquisition
  middleware, in: 2007 15th IEEE-NPSS Real-Time Conference, 2007, pp. 1--3.
\newblock \href {http://dx.doi.org/10.1109/RTC.2007.4382850}
  {\path{doi:10.1109/RTC.2007.4382850}}.

\bibitem{Blum:2008}
W.~Blum, W.~Riegler, L.~Rolandi, {Particle Detection with Drift Chambers},
  Springer-Verlag (2008) Sec. 4.4.2\href
  {http://dx.doi.org/10.1007/978-3-540-76684-1}
  {\path{doi:10.1007/978-3-540-76684-1}}.

\bibitem{ref:gpp}
H.~Schindler, \href{https://garfieldpp.web.cern.ch/garfieldpp/}{Garfield++}.
\newline\urlprefix\url{https://garfieldpp.web.cern.ch/garfieldpp/}

\bibitem{Sharma:1994rc}
A.~Sharma, F.~Sauli, {Low mass gas mixtures for drift chambers operation},
  Nucl. Instrum. Meth. A 350 (1994) 470--477.
\newblock \href {http://dx.doi.org/10.1016/0168-9002(94)91246-7}
  {\path{doi:10.1016/0168-9002(94)91246-7}}.

\bibitem{Bernardini:1994db}
P.~Bernardini, G.~Fiore, R.~Gerardi, F.~Grancagnolo, U.~von Hagel,
  F.~Monittola, V.~Nassisi, C.~Pinto, L.~Pastore, M.~Primavera, {Precise
  measurements of drift velocities in helium gas mixtures}, Nucl. Instrum.
  Meth. A 355 (1995) 428--433.
\newblock \href {http://dx.doi.org/10.1016/0168-9002(94)01144-3}
  {\path{doi:10.1016/0168-9002(94)01144-3}}.

\bibitem{Grab:1992ej}
C.~Grab, {Studies of helium gas mixtures as a drift chamber gas}, in:
  {ETHZ-IMP-PR-92-5, B Factories: the State of the Art in Accelerators,
  Detectors, and Physics}, 1992, pp. 0443--448.

\bibitem{Avanzini:2000ve}
C.~Avanzini, et~al., {Test of a small prototype of the KLOE drift chamber in
  magnetic field}, Nucl. Instrum. Meth. A 449 (2000) 237--247.
\newblock \href {http://dx.doi.org/10.1016/S0168-9002(99)01372-8}
  {\path{doi:10.1016/S0168-9002(99)01372-8}}.

\bibitem{Uno:1992rs}
S.~Uno, et~al., {Study of a drift chamber filled with a helium - ethane
  mixture}, Nucl. Instrum. Meth. A 330 (1993) 55--63.
\newblock \href {http://dx.doi.org/10.1016/0168-9002(93)91304-6}
  {\path{doi:10.1016/0168-9002(93)91304-6}}.

\bibitem{Apostolakis:2000yu}
J.~Apostolakis, S.~Giani, L.~Urban, M.~Maire, A.~Bagulya, V.~Grishin, {An
  implementation of ionisation energy loss in very thin absorbers for the
  GEANT4 simulation package}, Nucl. Instrum. Meth. A 453 (2000) 597--605.
\newblock \href {http://dx.doi.org/10.1016/S0168-9002(00)00457-5}
  {\path{doi:10.1016/S0168-9002(00)00457-5}}.

\bibitem{Tanabashi:2018oca}
M.~Tanabashi, et~al., {Review of Particle Physics}, Phys. Rev. D 98~(3) (2018)
  030001.
\newblock \href {http://dx.doi.org/10.1103/PhysRevD.98.030001}
  {\path{doi:10.1103/PhysRevD.98.030001}}.

\bibitem{Andryakov:1998yk}
A.~Andryakov, et~al., {dE/dx measurement in a He-based gas mixture}, Nucl.
  Instrum. Meth. A 409 (1998) 390--394.
\newblock \href {http://dx.doi.org/10.1016/S0168-9002(98)00106-5}
  {\path{doi:10.1016/S0168-9002(98)00106-5}}.

\bibitem{Allison:1980vw}
W.~Allison, J.~Cobb, {Relativistic Charged Particle Identification by Energy
  Loss}, Ann. Rev. Nucl. Part. Sci. 30 (1980) 253--298.
\newblock \href {http://dx.doi.org/10.1146/annurev.ns.30.120180.001345}
  {\path{doi:10.1146/annurev.ns.30.120180.001345}}.

\bibitem{Moritsu:2019uoy}
M.~Moritsu, et~al., {Commissioning of the Cylindrical Drift Chamber for the
  COMET experiment}, PoS EPS-HEP2019 (2020) 128.
\newblock \href {http://dx.doi.org/10.22323/1.364.0128}
  {\path{doi:10.22323/1.364.0128}}.

\end{thebibliography}

\end{document}